# Radiation of the Tunnel Electron on Secondary Center of Recombination

P. A. Golovinski[1,2], A. A. Drobyshev[1]

[1] *Voronezh State University of Architecture and Civil Engineering 394006 Voronezh, Russia*
[2] *Moscow Institute of Physics and Technology (State University) 141700 Moscow, Russia*
E-mail: goloviski@bk.ru

We discuss the photon emission that occurs due to the radiative recombination of an electron on a nearby center after tunneling ionization. The model of an active electron is used, and analytical solution to three-dimensional problem is obtained. The dependence of the photon emission from a distance between centers of ionization and a recombination, and also the electric field orientation are investigated. The formulas for probability of recombination radiation are derived.

## 1. Introduction

In theoretical and experimental investigation of interaction of a laser field with atoms, molecules and solid state the considerable attention is concentrated on dynamics of electrons and the harmonic generation, arising under activity of femto and attosecond laser pulses [1-3]. The majority of experimental results, known for the high harmonic generation, are compatible with the model of a rescattered electron, suggested by Corcum [4, 5], in which radiation is caused by the active electron, which starts a motion in the field of a laser wave after ionization. Such a process is related to returning of the electron possessing large energy to a parent atom. Observability of the process, for low-frequency laser field, tends to decrease, because of the electron package dispersion and a large recovery time that becomes great for its effective interaction with a parent centre [6].

At the same time, non-linear ionization of atoms and molecules with the subsequent acceleration of electrons in a monochromatic low-frequency laser field of high intensity can lead to recombination of electrons on other close centers of capture with radiation of photon. Radiation at the action of laser on bulk structures has revealed linear dependence of the characteristic radiated frequencies from the field strength [7]. It strikingly differs from quadratic dependence on electric strength of a laser field for maximum frequency of harmonic generation on individual atoms [8]. Investigation of the influence of additional centre of recombination [9, 10] on the harmonic generation has shown increasing of the peak frequency of harmonics due to decreasing of laser field frequency, and also with the growth a distance between centers. The process of tunneling electron radiation in a uniform electric field with transition to the next bound state is observed in quantum cascade lasers, where the linear frequency dependence of radiation frequency from electric strength [11] is observed too.

In the present work, electron radiation caused by tunnel ionization of quantum system in a constant or low-frequency laser field $F(t)$ with the subsequent recombination on additional centre is considered. In the strong light fields and low-frequency limit the non-linear ionization is determined by a small value of the Keldysh's parameter $\gamma = \omega\sqrt{2|E_i|}/F_0 << 1$ ($\omega$ is the characteristic frequency of a field, $E_i$ is an electron binding energy) [12]. In this limit the ionization process is quasi-static, and the main part of ionization occurs near the peak value of field strength $F_0$ [13]. The success of the given model is proved by application of the ADK formula [14] to description of non-linear ionization of atoms.

We consider electron radiation as three-step process [15, 16]: tunnel ionization, acceleration in an external field and a photorecombination with electron attachment to second centre. If the electron oscillation amplitude $F_0/\omega^2$ in a laser field much exceeds the distance $R$ between centers, the process is possible to consider as quasi-static. Further quantum mechanical calculation gives a detailed quantitative description of the phenomenon.

## 2. Model of an Active Electron

We shall discuss a problem of recombination for a tunneling electron is in a uniform electric field with strength $F$, basing on the method of propagation function [17]. In Fig. 1 the main steps are shown for the process, in which free electron is formed during tunnel ionization, and then, after acceleration between centers, is recombined on the second centre of interaction with radiation of quantum.

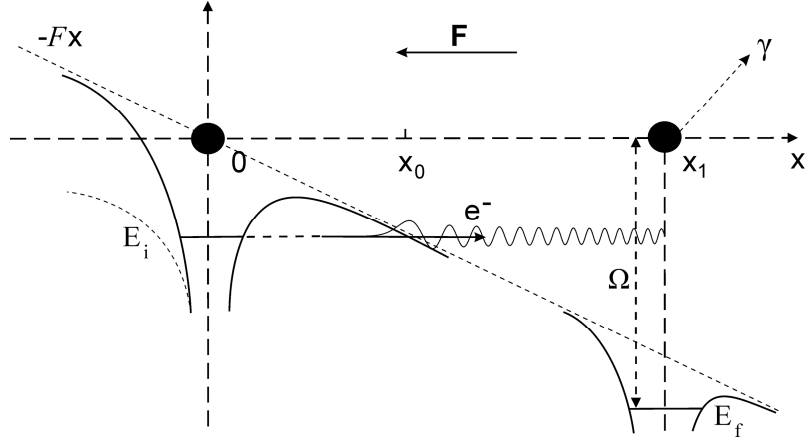

**Fig. 1.** Scheme of tunneling electron photorecombination

The Hamiltonian of a system has the form

$$H_t = H + W .\qquad(1)$$

Here the Hamiltonian $H$ of an electron incorporates the action of a strong laser field and perturbation [18]

$$W = -\left(\frac{2\pi}{V\Omega}\right)^{1/2} e^{-i\mathbf{Q}\mathbf{r}} (\mathbf{e}_\alpha \hat{\mathbf{p}}) e^{i\Omega t} \qquad(2)$$

is responsible for spontaneous radiation with frequency $\Omega$, polarization $\mathbf{e}_\alpha$, and a wave vector $\mathbf{Q}$ in uantization volume $V$. We adhere to an atomic units system, where $e = m = \hbar = 1$. Dealing with a process in a solid state the free electron mass should be changed by effective mass [19]. If we are restricted by dipole approximation $\exp(-i\mathbf{Q}\mathbf{r}) \approx 1$ and replace the operator of momentum by the operator in the length form $\hat{\mathbf{p}} \to -i\Omega\mathbf{r}$, we write down

$$W(\mathbf{r},t) = w(\mathbf{r})e^{i\Omega t}, w(\mathbf{r}) = i\left(\frac{2\pi\Omega}{V}\right)^{1/2} (\mathbf{e}_\alpha \mathbf{r}) .\qquad(3)$$

The Schrodinger equation with the specified Hamiltonian is of form

$$\left(i\frac{\partial}{\partial t} - H\right)\psi(\mathbf{r},t) = W\psi(\mathbf{r},t) .\qquad(4)$$

It is possible to express the solution to Eq. (4) through the exact Green's function $G$ of an electron [17] like

$$\psi(\mathbf{r},t) = i\int d^3r_1\, G(\mathbf{r},t;\mathbf{r}_1,t_1)\psi_i(\mathbf{r}_1,t_1), t > t_1, \qquad (5)$$

where $\psi_i(\mathbf{r}_1,t_1)$ is an electron wave function in an initial state. In the first order of a perturbation theory for the operator of spontaneous radiation $W$ the equation for the Green's function can be written down as

$$G(\mathbf{r},t;\mathbf{r}_1,t_1) = G_0(\mathbf{r},t;\mathbf{r}_1,t_1) + \int d^3r_2\, dt_2\, G_0(\mathbf{r},t;\mathbf{r}_2,t_2)W(\mathbf{r}_2,t_2)G_0(\mathbf{r}_2,t_2;\mathbf{r}_1,t_1). \qquad (6)$$

The Green's function $G_0(\mathbf{r},t;\mathbf{r}_1,t_1)$ includes the influence of external field on an electron motion.

The transition probability amplitude to the final bound state $\psi_f(\mathbf{r},t)$ is

$$S_{if} = \int d^3r\, \psi_f^*(\mathbf{r},t)\psi(\mathbf{r},t). \qquad (7)$$

If taking into account the Eq. (5), the Eq. (7) can be conversed to

$$S_{if} = i\int d^3r\, d^3r_1\, \psi_f^*(\mathbf{r},t)\, G(\mathbf{r},t;\mathbf{r}_1,t_1)\psi_i(\mathbf{r}_1,t_1). \qquad (8)$$

Using Eq. (6), we have

$$S_{if} = \int d^3r_2\, dt_2\, \varphi_f^*(\mathbf{r}_2)e^{i(\widetilde{E}_f+\Omega)t_2}\, w(\mathbf{r}_2)\psi_c(\mathbf{r},\ t). \qquad (9)$$

Here

$$\psi_c(\mathbf{r},\ t) = \int d^3r_1\, G_0(\mathbf{r}_2,t_2;\mathbf{r}_1,t_1)\psi_i(\mathbf{r}_1,t_1) \qquad (10)$$

is a wave function of active electron,

$$\psi_f(\mathbf{r},t) = \varphi_f(\mathbf{r})e^{i\widetilde{E}_f t} \qquad (11)$$

is a wave function of final state, and energy in this state is

$$\widetilde{E}_f = E_f - RF, \qquad (12)$$

with redefinition of energy due to the electric field potential variation between two centers (Fig. 1).

The most productive analytical approach featuring harmonic generation within the frame of isolated delta potential method is developed in [20-23]. In this model the wave function of an active electron $\psi_c(\mathbf{r},\ t) = \varphi_{E_i}(\mathbf{r})e^{-iE_i t}$ can be written down taking into account the energy conservation law in the uniform electric field. Within the theory of quasi-stationary states [18] one can take

$$\psi_c(\mathbf{r},\ t) = \int_0^\infty C(E)\varphi_E(\mathbf{r})e^{-iEt}\, dE, \qquad (13)$$

where

$$|C(E)|^2 = \frac{1}{\pi} \frac{\Gamma/2}{(E-E_i)^2 + (\Gamma/2)^2}, \qquad (14)$$

and $\Gamma$ is the width of quasi-stationary level equal to an imaginary part of quasienergy. Because

$$S_{if} = \int d^3x\, dt\, \varphi_f^*(\mathbf{r}) e^{i(E_f - RF + \Omega)t} w(\mathbf{r}) \psi_c(\mathbf{r},\, t), \qquad (15)$$

then, using Eq. (13), we have

$$S_{if} = \int_0^\infty C(E) \int_0^t dt_2\, e^{i(E_f - RF + \Omega - E)t_2} \int d^3r\, \varphi_f^*(\mathbf{r}) w(\mathbf{r}) \varphi_E(\mathbf{r})\, dE. \qquad (16)$$

Observable transition probability in unit of time $t$ for separate frequency $\Omega$ is proportional to quadrate of element of the position vector:

$$|S_{fi}|^2 / t = 2\pi |C(E)|^2 \left|\langle \varphi_f(\mathbf{r})|w(\mathbf{r})|\varphi_E(\mathbf{r})\rangle\right|^2. \qquad (17)$$

To calculate the transition probability in unit of time, one must multiply Eq. (17) by a number of possible final photon states $dN_f$. This number of states in volume $V$ at a certain polarization and momentum $p, p+dp$ in a spatial angle $do$ is given by expression

$$dN_f = \frac{V p^2 dp\, do}{c^2 (2\pi)^3}, \qquad (18)$$

and the distribution of states [18] over frequencies is equal to

$$d\rho(\Omega) = \frac{dN_f}{d\Omega} = \frac{V \Omega^2\, do}{(2\pi c)^3}. \qquad (19)$$

Taking into account Eq. (19), the probability density of quantum emission with frequency $\Omega$, polarization $\mathbf{e}_\alpha$ in a spatial angle $do$ we write down as

$$dP/d\Omega = \frac{\Omega^3}{2\pi c^3} |C(E)|^2 |\mathbf{D}_\Omega|^2 \sin^2\theta\, do, \qquad (20)$$

and matrix element of a position vector

$$\mathbf{D}_\Omega = \langle \varphi_f(\mathbf{r})|\mathbf{r}|\varphi_E(\mathbf{r})\rangle. \qquad (21)$$

Variable $\theta$ is an angle between direction of the vector $\mathbf{D}_\Omega$ and wave vector $\mathbf{Q}$. Integration over spatial angle leads to

$$\frac{dP}{d\Omega} = \frac{4\Omega^3}{3c^3} |C(E)|^2 |\mathbf{D}_\Omega|^2. \qquad (22)$$

Then the distribution of intensity over frequencies [24] is of form

$$\frac{dI_\Omega}{d\Omega} = \frac{4\Omega^4}{3c^3}|C(E)|^2|\mathbf{D}_\Omega|^2. \tag{23}$$

To calculate the total intensity of a line, the shape factor needs to be changed by integral over energy, and that is equal to unit.

### 4. Quantum Dots

For delta-potential model [35], the potential has the form

$$V(r) = -V\delta(\mathbf{r})\frac{\partial}{\partial r}r. \tag{24}$$

The equation for the electron wave function in the presence of a uniform electric field is

$$\left(E_i + \frac{1}{2}\nabla^2 + Fx\right)\varphi(\mathbf{r}) = \frac{2\pi}{\kappa}\delta(\mathbf{r})\left(\frac{\partial}{\partial r}r\varphi(\mathbf{r})\right). \tag{25}$$

For $F = 0$ the bound state [25] is of form

$$\varphi_0(r) = \sqrt{\frac{\kappa}{2\pi}}\frac{\exp(-\kappa r)}{r} \tag{26}$$

with a binding energy $E_i = -\kappa^2/2$. The equivalent integral equation for the wave function is

$$\varphi(\mathbf{r}) = \frac{2\pi}{\kappa}\int d^3r_1\, G_E(\mathbf{r},\mathbf{r}_1)\delta(\mathbf{r}_1)\left(\frac{\partial}{\partial r_1}r_1\varphi(\mathbf{r}_1)\right). \tag{27}$$

Here $G_E(\mathbf{r},\mathbf{r}_1)$ is the Green's function. By substituting the unperturbed wave function of Eq. (26) at the right hand side of Eq. (27), we obtain

$$\varphi(\mathbf{r}) = -\sqrt{\kappa 2\pi}\,G_E(\mathbf{r},0). \tag{28}$$

Matrix element of a position vector is equal

$$\mathbf{D}_\Omega = -\sqrt{\kappa 2\pi}\langle\varphi_f(\mathbf{r})|\mathbf{r}|G_E(\mathbf{r},0)\rangle, \tag{29}$$

where

$$G_E(\mathbf{r},0) = \frac{1}{(2\pi)^2}\iint \exp(ip_y y + ip_z z)G_{E-p^2/2}(x,0)dp_y dp_z. \tag{30}$$

The one-dimensional Green's function $G(x,x_1,E)$ has the form [26]:

$$G(x, x_1, E) = -\frac{2\pi}{a} \text{Ci}(-ax_> + b) \text{Ai}(-ax_< + b). \tag{31}$$

In the Eq. (31) we used notation

$$a = (2F)^{1/3}, \quad b = -2E/a^2 \tag{32}$$

and

$$x_> = \frac{1}{2}(x + x_1 + |x - x_1|), \quad x_< = \frac{1}{2}(x + x_1 - |x - x_1|). \tag{33}$$

Standard Airy functions are denoted $\text{Ai}(x)$, $\text{Bi}(x)$ [27], and

$$\text{Ci}(x) = \text{Bi}(x) + i\text{Ai}(x). \tag{34}$$

The function $\text{Ci}(x)$ at $x \to \infty$ has asymptotic behavior of an outgoing wave. Substitution of Eq. (30) to Eq. (29) gives an expression

$$\mathbf{D}_\Omega = -\frac{\sqrt{\kappa 2\pi}}{(2\pi)^2} \int \langle \varphi_f(\mathbf{r}) | \mathbf{r} | G_{E-p^2/2}(x, 0) \exp(i(p_y y + p_z z)) \rangle dp_y dp_z. \tag{35}$$

The final state we take in the approximation of three-dimensional linear oscillator [28]

$$\varphi_f(\mathbf{r}) = \varphi_f(x)\varphi_f(y)\varphi_f(z) = \varphi_0(x - x_1)\varphi_0(y - y_1)\varphi_0(z - z_1), \tag{36}$$

$$\varphi_0(x) = (\omega_0/\pi)^{1/2} \exp(-\omega_0 x^2/2).$$

where $x_1, y_1, z_1$ are coordinates of recombination centre. If the axis OX is directed along the motion of a free electron and through ionization center and the plane OXY contains recombination centre, then $z_1 = 0$. Taking this into account, one obtains

$$\mathbf{D}_\Omega = -\frac{\sqrt{\kappa 2\pi}}{(2\pi)^2} \int \langle \varphi_0(x - x_1)\varphi_0(y - y_1)\varphi_0(z) | \mathbf{r} | G_{E-p^2/2}(x, 0) \exp(i(p_y y + p_z z)) \rangle dp_y dp_z. \tag{37}$$

This formula allows one to calculate photorecombination for three-dimensional problem.
In the Fig. 2 the spectrum of recombination on a quantum dot is depicted. In the Fig. 3 the dependence of the recombination probability from distance $y_1$ (from the centre to the axis OX) is depicted. By displacement of recombination centre along perpendicular to the electric field a rapid decrease in the intensity of radiation can be obtained. Distribution of an electronic current has approximate expression $j_z \sim \exp(-\kappa\rho^2/2x)$ in a transverse direction, and it determines the characteristic distance $y_1 \sim \sqrt{2x/\kappa}$ on which the recombination is terminated.

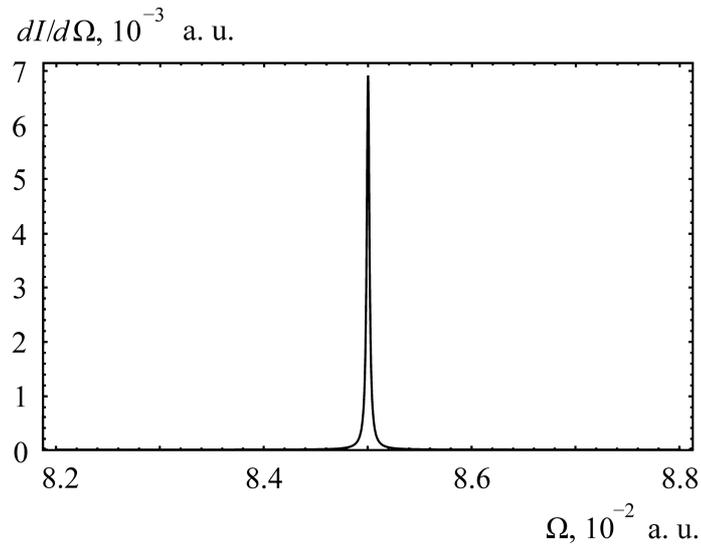

**Fig. 2.** The spectrum of radiation for $y_1 = 10$ a.u. The distance from centers $R = 50$ a. u., $z_1 = 0$ a. u., $I = 10^{11}$ W/cm$^2$, $E_i = E_f = 1$ eV.

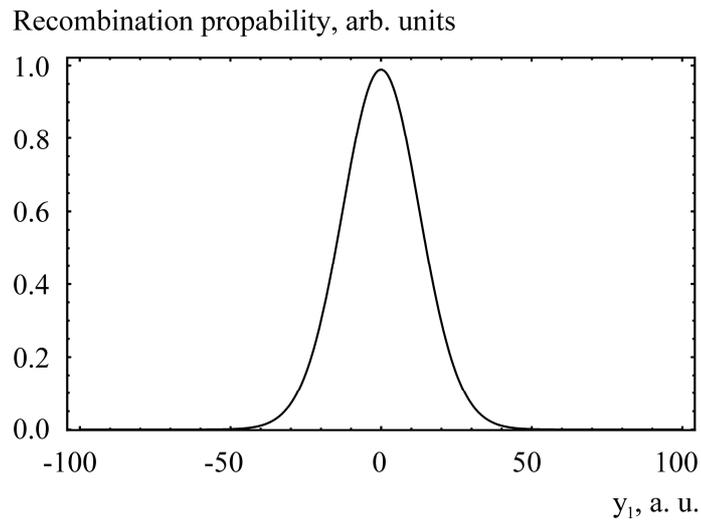

**Fig. 3.** The dependence of recombination probability as a function of $y_1$. The distance from centers $R = 50$ a. u., $z_1 = 0$ a. u., $I = 10^{11}$ W/cm$^2$, $E_i = E_f = 1$ eV.

### 4. Conclusions

The phenomenon of a photorecombination of a tunnel electron in a uniform electric field, considered for the case of second close located centre, has some properties that allow to identify this process with respect to other radiative processes. Radiated frequency in such process is in proportion to the electric field strength and to the distance between centers of ionization and recombination. This dependence originates quite naturally, and it is related to acceleration of electron in a gap between its exit under a field barrier and quantum emission. The difference with a classical behavior at a small distance between centers is a result of dispersion of a wave function after electron exit from under a barrier. In experimental studies of the photorecombination, following the tunnel ionization, it is convenient to observe the dependence of recombination radiation as a function of electric field strength and polarization orientation.

The work was supported by RFFR (the grant 13-07-00270) and by RF government contract № 14.513.11.0133.

**References**

1. I. J. Sola, E. Mevel, L. Elouga, et. al., Nature Phys. **2**, 319 (2006).
2. P. B. Corcum, F. Krausz, Nature Phys. **3**, 381 (2007).
3. E. A. Volkova, A. M. Priests, O .V. Tikhonov, Letters in ZhETF **94**, 559 (2011).
4. P. B. Corkum, Phys. Rev. Lett. **71**, 1994 (1993).
5. K. C. Kulander, J. Cooper, K. J. Schafer, Phys. Rev. A **51**, 561 (1995).
6. P. A. Golovinski, Laser Phys. **3**, 280 (1993).
7. S. Ghimire, A. D. DiChiara, E. Sistrunk, P. Agostini, et. al, Nature Phys. **7**, 138 (2011).
8. Th. Brabec, F. Krausz, Rev. Mod. Phys. **72**. 545 (2000).
9. R. Kopold, W. Becker, Laser Phys. **9**, 398 (1999).
10. Y. Wang, Zh. Sun X. Zhang et. al., Chinese Opt. Lett. **4**, 49 (2006).
11. M. Beck, D. Hofstetter, Th. Aellen et. al., Science **295**, 301 (2002).
12. A. I. Baz, Ja. B. Zel'dovich, A. M. Perelomov. Reactions, scattering and decays in the nonrelativistic quantum mechanics, Moscow, Nauka, 1971 (in Russian).
13. N. B. Delone, V. P. Krainov. Non-linear ionisation of atoms by a laser radiation, Moscow, Fizmatlit, 2001 (in Russian).
14. M. V. Ammosov, N. B. Delone, V. P. Krainov, ZhETF **91**, 2008 (1986).
15. S. L. Chin, P. A. Golovinski, J. Phys. B.: At. Mol. Opt. Phys. **28**, 55 (1995).
16. M. Yu. Kuchiev V. N. Ostrovsky, Phys. Rev. A **60**, 3111 (1999).
17. J. D. Bjorken, S. D. Drell. Relativistic Quantum Theory, V.1, NY, MGH, (1964).
18. A. S. Davidov. Quantum mechanics, SPb: BVH-Peterburg, 2011 (in Russian).
19. H. Haug, S. W. Koch. Quantum theory of the optical and electronic properties of semiconductors, Singapore: World Scientific, 2004.
20. W. Becker, S. Long and J. K. McIver, Phys. Rev. A. **50**, 1540 (1994).
21. M. V. Frolov, N. L. Manakov, T. S. Sarantseva, et. al., J. Phys. B.: At. Mol. Opt. Phys. **42**, 035601 (2009).
22. M. V. Frolov, N. L. Manakov, A. F. Starace, Phys. Rev. A. **78**, 063418, 2008.
23. M. V. Frolov, N. L. Manakov, T. S. Sarantseva, et. al., J. Phys. B.: At. Mol. Opt. Phys. **42**, 035601 (2009).
24. Yu. N. Demkov, V. N. Ostrovsky. A method of potentials with zero radius in the atomic physics, L, LSU, 1975 (in Russian).
25. Yu. N. Demkov, G. F. Drukarev, ZhETF **47**, 918, (1964)
26. B. Gottlieb, M. Kleber, J. Krause, Z. Phys. A. **339**, 201 (1991).
27. M. Abramowitz, I. A. Stegun. Handbook of mathematical functions, Washington, National Bureau of Standards, 1972.
28. L. D. Landau, E. M. Lifshits. Quantum mechanics, Moscow, Fizmatlit, 2008 (in Russian).